\documentclass[a4paper,12pt]{spieman}
\usepackage{amsmath,amsfonts,amssymb}
\usepackage{graphicx}
\usepackage{setspace}
\usepackage{tocloft}
\usepackage[nolist]{acronym}
\newcommand{\resetnoiserms}{\ensuremath{\sigma_{\mathrm{rms}} \left[\mathrm{e^-}\right]}}

\begin{acronym}
	\acro{ADC}{Analog to Digital Converter}
	\acro{CCD}{Charge Coupled Device}
	\acro{CDS}{Correlated Double Sampling}
	\acro{DA}{Differential Averager}
	\acro{DCDS}{Digital Correlated Double Sampling}
	\acro{DSI}{Dual Slope Integrator}
	\acro{LR}{Linearity Residual}
	\acro{LSST}{Large Synoptic Survey Telescope}
	\acro{PTC}{Photon Transfer Curve}
	\acro{SNR}{Signal to Noise Ratio}
\end{acronym}

\title{Automatic Selection of \ac{CDS} Timing Parameters}

\author[a]{D. P. Weatherill}
\author[a]{I. Shipsey}
\author[a]{K. Arndt}
\author[a]{R. Plackett}
\author[a]{D. Wood}
\author[a]{K. Metodiev}
\author[a]{M. Mironova}
\author[a]{D. Bortoletto} 
\author[a]{N. Demetriou}
\affil[a]{University of Oxford, Department of Physics, Keble Road, Oxford, UK OX1 3RH}

\begin{document}
	\maketitle
	
	\begin{abstract}
		 \ac{CDS} is a process used in many \ac{CCD} readout systems to cancel the reset noise component that 
		 would otherwise dominate. CDS processing typically consists of subtracting the integrated video 
		 signal during a "signal" period from that during a "reset" period. The response of this processing 
		 depends therefore on the shape of the video signal with respect to the integration bounds. In 
		 particular, the amount of noise appearing in the final image and the linearity of the pixel 
		 value with signal charge are affected by the choice of the CDS timing intervals. 

		 In this paper, we use a digital \ac{CDS} readout system which highly oversamples the video signal 
		 (as compared with the pixel rate) to reconstruct pixel values for different CDS timings using 
		 identical raw video signal data. We use this technique to develop insights into optimal strategy for 
		 selecting CDS timings both in the digital case (where the raw video signal may be available), and in 
		 the general case where it is not.
		 
		 In particular, we show that the linearity of the CDS operation allows subtraction of the raw video signals of pixels in bias images from those in illuminated images to directly show the effects of CDS processing on the final (subtracted) pixel values.
	\end{abstract}
	\keywords{Charge Coupled Devices, Correlated Double Sampling, Readout Electronics, Astronomy}
	\section{Introduction}
	Modern \acp{CCD} are capable of extremely low read noise which is crucial to the science 
	goals of current and future astronomical observatories. For example, the baseline requirement of the 
	sensors to be used in the \ac{LSST} camera is a readout noise of 5 e$^-$ rms equivalent at a pixel 
	frequency of 500 kpix s$^{-1}$ \cite{1748-0221-4-03-P03002}. This requirement is driven by the need to 
	maintain the instrumental readout noise below the sky background shot noise levels, in particular in the 
	short wavelength u band \cite{2008arXiv0805.2366I}.
	
    	Typically the dominant on chip noise source of a \ac{CCD} is the reset noise, which arises from the Johnson-Nyquist noise associated 
    	with the resetting of the sense node used to convert the accumulated charge into a voltage 
    	signal\cite{janesick2001scientific}. In order to achieve the best possible performance, this noise 
    	signal is typically removed via a process known as \ac{CDS}. The remaining readout noise is then 
    	associated with thermal effects, carrier trapping and semiconductor parameter variation within the 
    	amplification transistors (so-called "read noise"), which for a given bandwidth represents the lowest 
    	noise achievable using a conventional \ac{CCD} output design.
	The \ac{CDS} process can be implemented using various types of analog \cite{0022-3735-15-11-020} or 
	digital \cite{oro42476} methods, though in all cases the timing parameters of the \ac{CDS} circuit 
	operation must be correctly chosen to properly eliminate reset noise without introducing adverse 
	artefacts into the measured pixel values in the form of structured noise patterns or excess non-linearity.
	
	In this paper we present investigations into the effects of these timing 
	parameters using a digital 
	\ac{CDS} system and an e2v CCD250 sensor \cite{doi:10.1117/12.2069423}. We perform analysis of the raw 
	oversampled video data (before processing to produce pixel values) to find correct timing parameters for 
	linearity and \ac{SNR} using numerical optimisation (see Section \ref{sec:optimising}). Using the 
	linearity property of the digital \ac{CDS} operation, we also show that performing subtraction of the raw 
	video signal for different pixels can give insight into the choice of timing parameters (see Section \ref{sec:subtraction}), a process which 
	might usefully be implemented on any \ac{CDS} system, even those 
	without access to oversampled raw video data.
	
	\section {Correlated Double Sampling Background}
	\begin{figure}
		\begin{center}
			\includegraphics{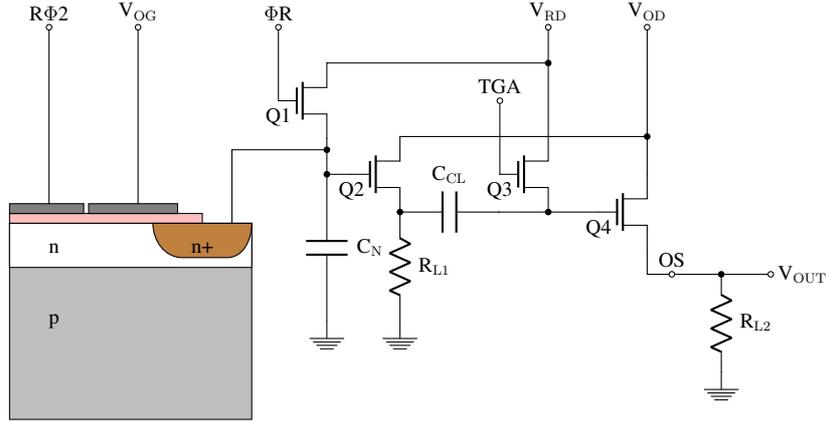}
		\end{center}
		\caption{\label{fig:output_schematic} Schematic representation of a 2-stage CCD output. 
			All components shown except the second stage load $\mathrm{R_{L2}}$ are on-chip. 
			The capacitor marked $\mathrm{C_N}$ represents the junction capacitance of the pn junction sense 
			node.}
	\end{figure}
	
\begin{figure}
			\begin{center}
			\includegraphics[]{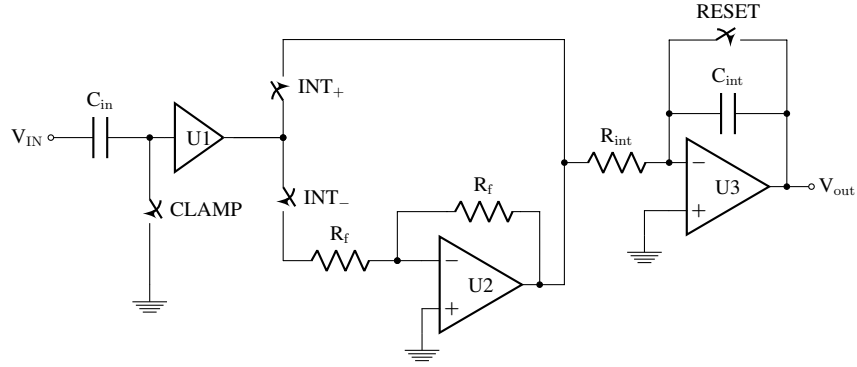}
			\end{center}
			\caption{\label{fig:dsi_schematic} Schematic representation of a \ac{DSI} circuit. }
	\end{figure}

	A diagram of a typical \ac{CCD} two stage on-chip amplifier circuit is shown in Figure 
	\ref{fig:output_schematic}. Reset noise is incurred in the readout process during the operation of the 
	sense node reset transistor (labelled Q1 in Figure \ref{fig:output_schematic}). When the reset signal 
	$\mathrm{\Phi R}$ is asserted, the voltage on the sense node is allowed to settle to the reset drain 
	voltage $\mathrm{V_{RD}}$. If the channel resistance of the reset transistor is $\mathrm{R_{Q1}}$, there 
	is a Johnson-Nyquist noise $\sigma_{\mathrm{rms}}$ on the voltage given by \cite{PhysRev.32.110}:
	\begin{equation} \label{reset_noise}
		\sigma_{\mathrm{rms}}=2\sqrt{k_BTB \mathrm{R_{Q1}} } 
	\end{equation}
	
	where $T$ is the temperature, $k_B$ is Boltzmann's constant, and $B$ is the bandwidth of the circuit. 
	This bandwidth is set by the RC circuit formed by the finite channel resistance of the reset transistor 
	$\mathrm{R_{Q1}}$ and the sense node capacitance $\mathrm{C_N}$. Using the transfer function of an RC 
	filter we obtain the amount of noise present on the reset $\sigma_{\mathrm{rms}}$ expressed in units of 
	electrons
	\begin{equation}
		\resetnoiserms = \frac{\sqrt{k_BT\mathrm{C_N}}}{q_e}
	\end{equation}
	
	where $q_e$ is the electronic charge. For the nominal \ac{LSST} operating temperature of $T = 203
	\mathrm{K}$  and assuming a typical sense node capacitance of $\mathrm{C_N} = 15 \ \mathrm{fF}$ leads to 
	 $\resetnoiserms \approx 40 \mathrm{e^-} $.	
	The reset noise can be effectively cancelled by measuring the individual reset level for each pixel 
	readout and subtracting the signal level from it using a \ac{CDS} process. Perhaps the most common 
	implementation of \ac{CDS} in the analog domain is the \ac{DSI}, also known as the \ac{DA}, which is 
	illustrated in Figure ~\ref{fig:dsi_schematic}. It is useful to refer to this diagram 
	even in the \ac{DCDS} case, since the operation of the digital \ac{DA} system is equivalent to this 
	circuit in the limit of infinite sample rate.  In fact, taking some simplifying 
	assumptions, it can be shown that a \ac{DSI} is a matched filter for the subtraction of pixel values in 
	the absence of $\frac{1}{f}$ noise\cite{oro42476}. Optimal filtering can also be designed in the presence 
	of $\frac{1}{f}$ noise, using either advanced digital filtering methods enabled by a \ac{DCDS} 
	system\cite{1538-3873-115-811-1068} or modified analog clamp \& sample circuitry\cite{101279}. However, 
	the question of selecting the best timing parameters for the circuit remains, even where the gain and (if 
	applicable) filter coefficients have been matched to the system noise spectrum.
	
\begin{figure}
		\centering
		\includegraphics{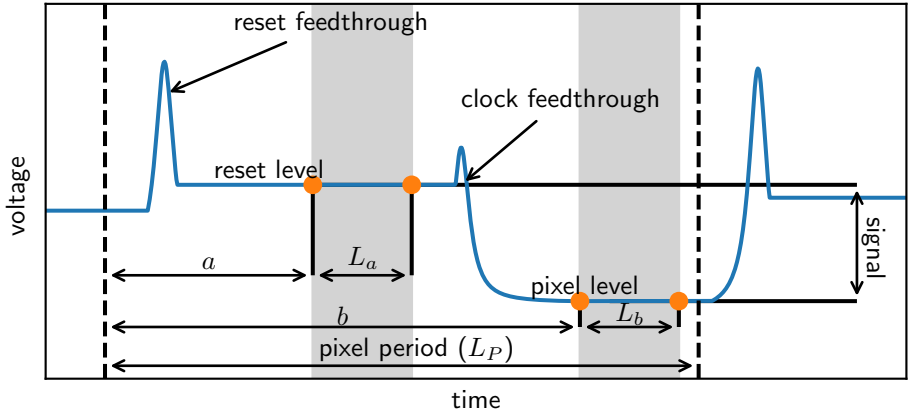}
		\caption{\label{fig:video_signal} Cartoon diagram of a \ac{CCD} video signal output. $(a,b,L_a,L_b)$ are the CDS timing parameters. $L_p$ is the pixel period}
	\end{figure}

	A diagram of how a video signal looks before \ac{CDS} processing is shown in Figure 
	\ref{fig:video_signal}. Throughout this work we label four \ac{CDS} timing parameters:
	\begin{itemize}
		\item $a$ - the offset from the start of the pixel to the beginning of the reset sampling window
		\item $L_a$ - the length of the reset sampling window
		\item $b$ - the offset from the start of the pixel to the beginning of the signal sampling window
		\item $L_b$ - the length of the signal sampling window
	\end{itemize}
	A processed pixel value is obtained by a simple procedure, described here in terms of the operation of 
	the circuit shown in Figure ~\ref{fig:dsi_schematic}. The CLAMP and RESET switches are typically operated 
	simultaneously with the reset feedthrough transient, to restore the DC level of the processor. At time 
	$a$, the switch $\mathrm{INT_-}$ is closed, and remains so for a period $L_a$, during which the negative 
	integral of the video reset window accumulates on the capacitor $C_\mathrm{int}$. At time $b$, switch 
	$\mathrm{INT_+}$ is closed, which causes the integral of the signal window to be added to that of the 
	reset window. The resulting output is the pixel value. The procedure for \ac{DCDS} is conceptually the 
	same, except that the signal is oversampled during the sampling windows by a fast \ac{ADC}, and the 
	integration and subtraction is performed digitally. 
	
	\section{Methods}
	The test system consists of an Teledyne-e2v CCD250 device cooled via liquid nitrogen to an operating 
	temperature 
	of $-100 ^\circ \mathrm{C}$, read out using a STA Archon system\cite{doi:10.1117/12.2058402}, and 
	illuminated by a stabilised Quartz-Tungsten light source passed through a monochromator. A more detailed 
	description of this test system has previously been published\cite{Weatherill_2017}.
	
	The Archon readout carries 16 16-bit 100MHz \acp{ADC}, one for each output channel of the CCD250. The 
	\ac{CDS} is performed internally to the controller, and it is not possible to apply non-unity weighting 
	co-efficients to the samples before processing. In this manner the system quite closely approximates a 
	\ac{DSI}, even for the fairly rapid pixel rate used. It is possible to read out the raw (pre-\ac{CDS}) 
	sample values for a specified region of an image for a single channel at a time. For all the 
	results shown here, we selected a region of 512 x 200 pixels (each channel has 512 columns and 2002 rows 
	in total). This results in approximately 220MB of raw data per captured image. The timing sequence used  
	yielded a pixel rate of 490 kHz, resulting in a pixel period $L_P = 204$  (all \ac{CDS} 
	timing parameters are given 
	in numbers of 100 MHz samples).
	
	Flat-field data was captured for 8 of the 16 total channels, each 
	consisting of 5 bias frames and 40 
	pairs of illuminated frames up to an integration time of 5 s. The backside bias of the \ac{CCD} was set 
	to $V_B = -60 \mathrm{V}$, and in future it would be of interest to study the effect of changing bias 
	voltages (the backside bias used significantly alters readout conditions, including the capacitance of 
	the sense node\cite{doi:10.1117/12.876627}). Using the raw sample data, we can then \emph{post facto} 
	reconstruct a \ac{PTC} for a set of \ac{CDS} timing parameters in software using exactly the same 
	underlying readout data. This process is very simple and consists of summing and normalising the sample 
	values in an identical manner to the \ac{DA} in the Archon firmware. We verified for two randomly 
	selected sets of parameters that our reconstructed pixel values were identical to the Archon pixel values 
	with \ac{CDS} enabled. We construct the \ac{PTC} following the standard procedure, including subtraction 
	of a mean overscan value for each row, subtraction of a bias frame before calculation of mean value, and 
	differencing of two illuminated images to eliminate fixed pattern noise before calculating 
	variance\cite{janesick2007dn}.
	
	\section{The Effect of CDS Timing on Linearity and Noise \label{sec:ptc_lin_noise}}
	\begin{figure}
		\centering
		\includegraphics{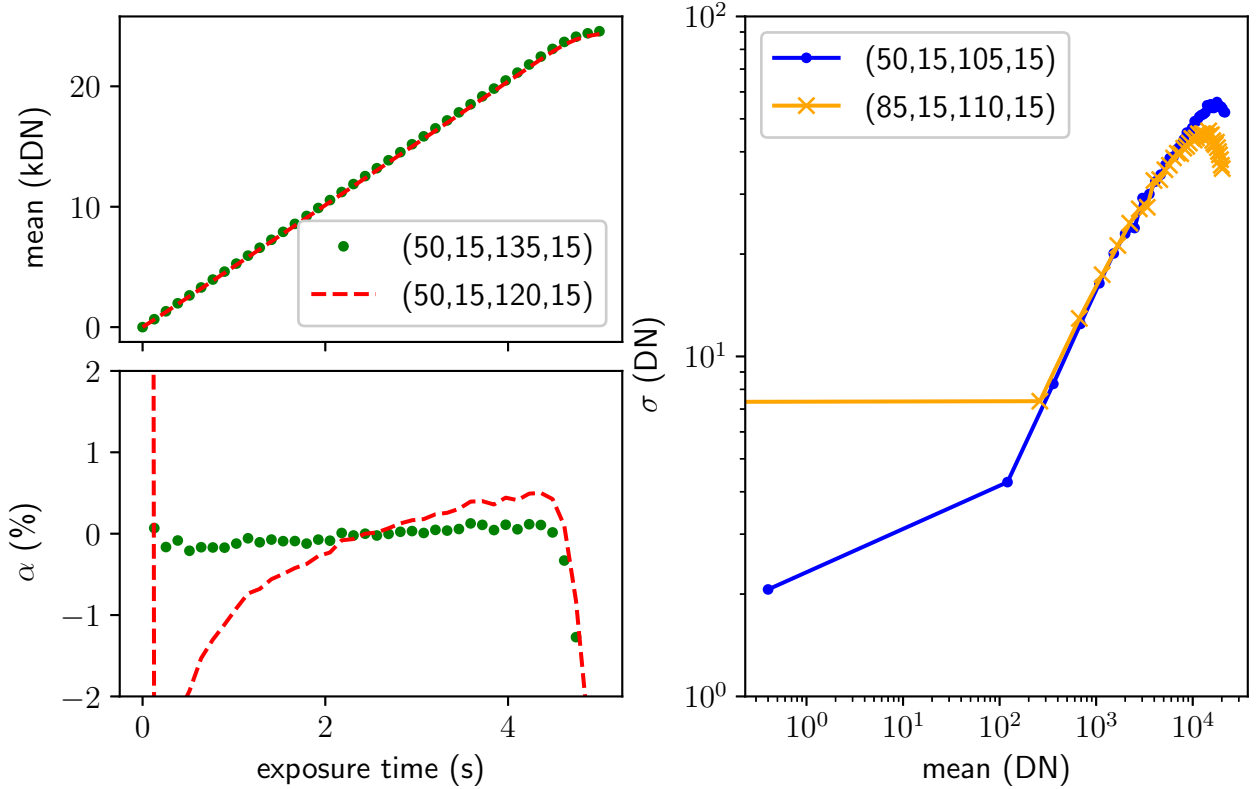}
		\caption{\label{fig:PTC_noise} Examples of PTC (right panel) and linearity (left panels) curves for 
		different CDS timing values. The timing parameters are given as $(a,L_a,b,L_b)$}

	\end{figure}
	It is clear that the values of $L_a$ and $L_b$ significantly affect the \ac{SNR} of the image. Intuitively, longer sampling times suppress the white noise component. This is found to be the case in practice. The values of $a$ and $b$ also affect \ac{SNR} because sampling near the region of a clock edge transition (which injects noise) increases the total integrated noise. This second 
	effect is illustrated in Figure \ref{fig:PTC_noise} (right panel) where the change of $(a,b) = (50,105)$ to $(a,b) = (85,110)$ significantly increases the 
	measured noise floor $\sigma$. 
	
	We also calculate the \acp{LR} $\alpha$ for the data according to the formula from 
	Janesick\cite{janesick2001scientific}:
	\begin{equation}
		\alpha_i = 100 \times \left(1 - \frac{S_\mathrm{mid} t_i}{S_i t_\mathrm{mid}}\right)
	\end{equation}
	where $S_i$ and $t_i$ are the mean signal levels and integration times respectively; and $S_\mathrm{mid}$ 
	and $t_\mathrm{mid}$ are the mean signal level and integration time at a selected midpoint. In this work 
	we use $t_\mathrm{mid} = 2.5 \mathrm{s}$. Again, it is intuitive that having sampling periods which 
	either overlap	a clock feedthrough or are too early after the decay of the reset transient will incur 
	significant linearity errors, since the magnitudes of these transients do not depend linearly on the 
	sense node charge. In Figure \ref{fig:PTC_noise} we see this in action. We start from a very poor 
	linearity situation ( $\left|\alpha_i\right|_\mathrm{max} > 2 \%$) with $(a,b) = (50,120)$ and 
	dramatically improve by making the signal sampling period later with $(a,b) = (50,135)$. We expect, 
	however, that $L_a$ and $L_b$ should have weak effect on linearity by themselves, though too large a 
	choice for these parameters would force the sampling periods into problematic regions. Thus the set of 
	timing parameters which optimise \ac{SNR} is almost always not the same as that which optimises \ac{LR}, 
	and that these two goals are in contention. It is also conceivable that a different set of timing 
	parameters would optimise linearity and \ac{SNR} over some signal ranges than others.

	\section{Optimising CDS parameters \label{sec:optimising}}
	Strictly, the optimisation problem presented to us is one of mixed-integer programming (since the timing 
	values are restricted to be integers), which is well known to be NP-hard. 
	
	However, through analogy to an analog \ac{CDS} system where the 
	timing parameter values are continuous, we expect that were it	possible to select non-integer sample 
	numbers, all the resulting cost functions should be well defined 
	and smooth. Hence, it is suitable to use standard Nelder-Mead multi-variate optimisation routines 
	(as implemented in the scipy package)\cite{scipy}, and to calculate the cost functions for non-integer 
	values by linearly interpolating from the nearest integer value results. As mentioned in Section 
	\ref{sec:ptc_lin_noise}, it is likely that for some applications (e.g low light imaging) one might tailor 
	a cost function to optimise \ac{CDS} timing in some specified signal range, or to emphasize linearity and 
	\ac{SNR} to different degrees. In this work, we consider the optimisation over the following cost 
	functions:
	\begin{align}
		f_\mathrm{SNR} = \frac{\left<S\right>_\mathrm{pix}}{\sigma_{\mathrm{bias}}} \\
		f_\mathrm{lin} = \sqrt{\left<\alpha^2\right>} \\
		f_\mathrm{comb} = \frac{f_\mathrm{SNR}}{f_\mathrm{lin}}
	\end{align}
	
	where $\left<x\right>$ indicates taking a mean average. Several more natural cases worthy of 	
	investigation would be the maximum \ac{LR} (as opposed to the rms value represented by $f_\mathrm{lin}$), 
	and combining \ac{SNR} and linearity with different powers in $f_\mathrm{comb}$, which are not considered 
	further here.
	\begin{figure} 
		\includegraphics{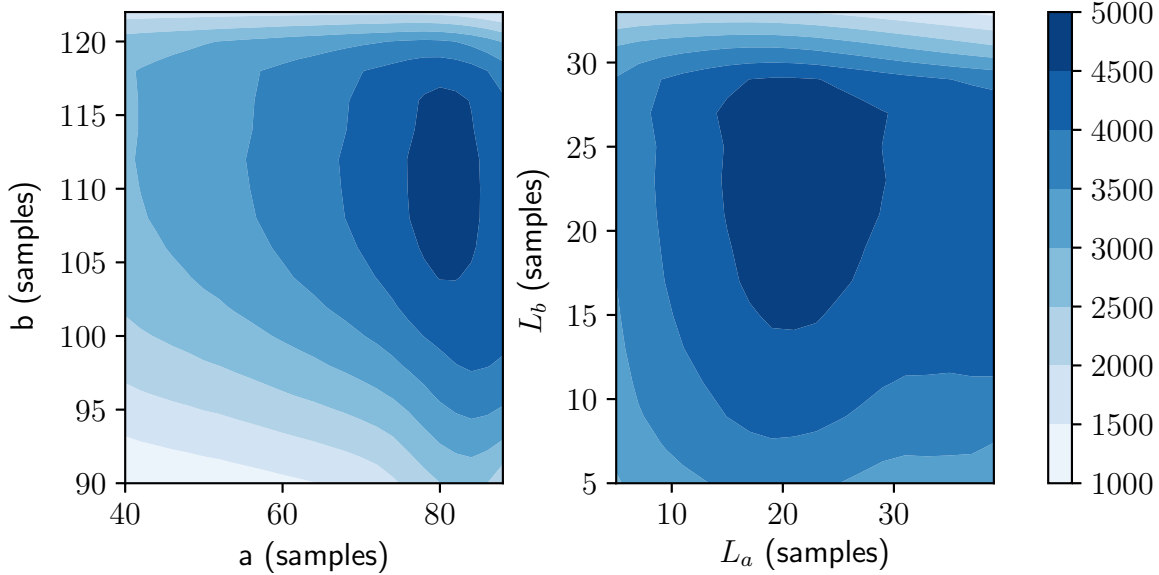}
		\caption{\label{fig:fSNR_sections} Cross sections for $f_\mathrm{SNR}$ through $(a,b)$ at constant 
		$L_a=L_b=20$ (left panel) and through $(L_a,L_b)$ at $(a,b) = (80,110)$ (right panel)}
	
	\end{figure}
	
\begin{figure} 
		\includegraphics[]{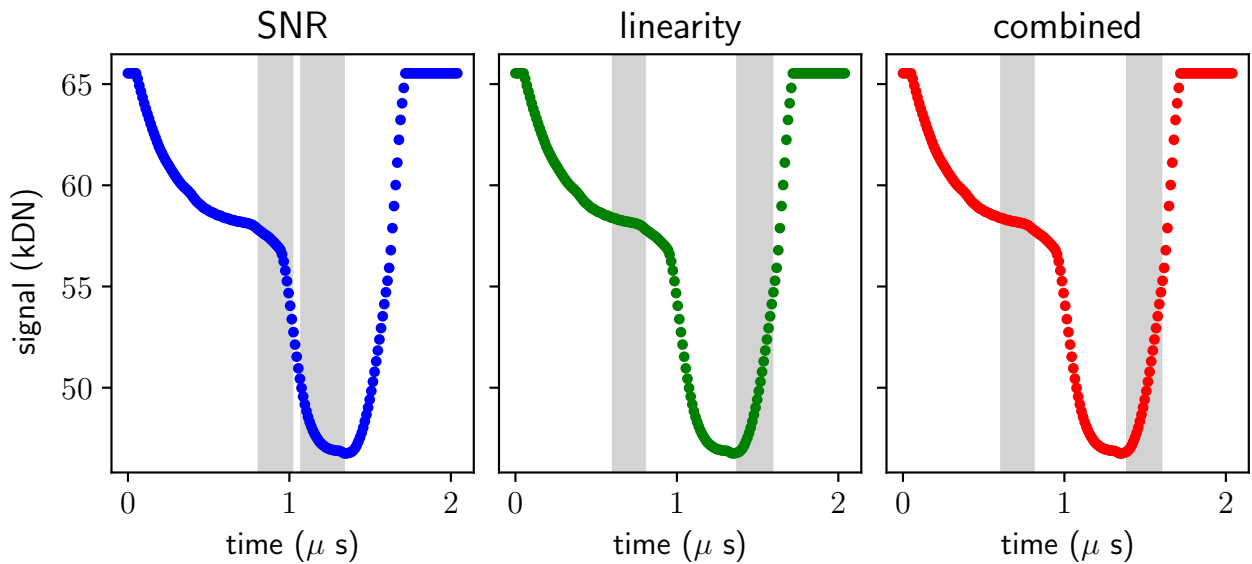}
		\caption{\label{fig:optim_CDS} Optimum CDS timing parameters (grey shaded) and corresponding example 
		raw pixel values}
	\end{figure}

	Two cross sections for $f_\mathrm{SNR}$ in $(a,b)$ and $(L_a, L_b)$ are shown in Figure 
	\ref{fig:fSNR_sections}. These give a somewhat intuitive picture, with a longer and later reset period 
	$L_a$ clearly improving \ac{SNR} to avoid the reset transient. The resulting parameters from the full 
	4-dimensional maximisation is shown in Figure \ref{fig:optim_CDS} (left panel). The optimised parameters 
	appear to contain significant parts of the clock feedthrough region, which will clearly be detrimental to 
	linearity performance.
	\begin{figure} 
		\includegraphics[]{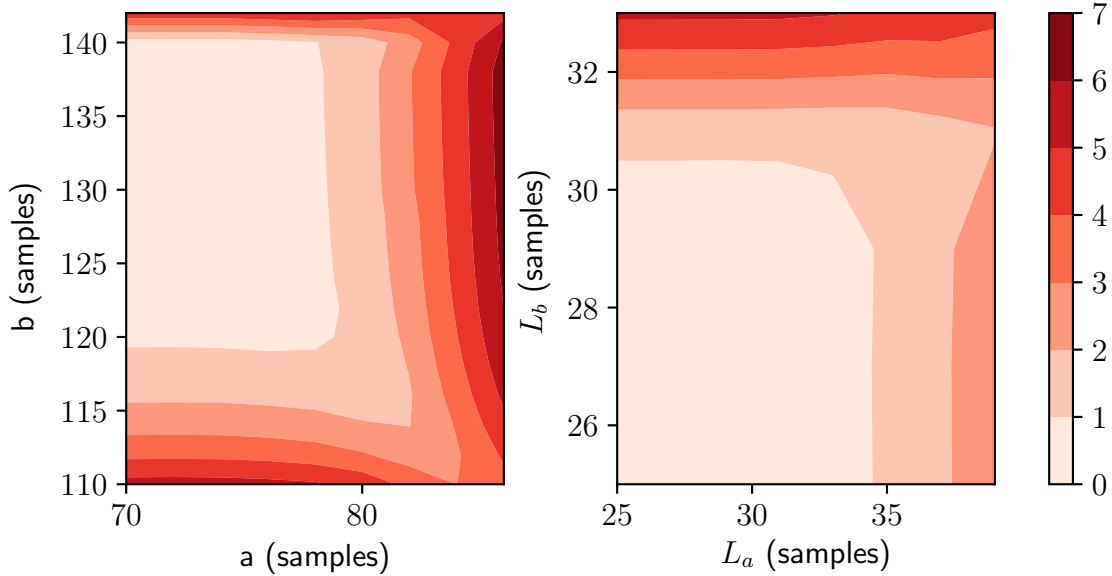}
		\caption{\label{fig:flin_sections} Cross sections for $f_\mathrm{lin}$ through $(a,b)$ at constant 
		$L_a=L_b=20$ (left panel) and through $(L_a,L_b)$ at $(a,b) = (75,130)$}
	\end{figure}
	
	Next the minimisation of $f_\mathrm{lin}$ is considered. Cross sections in $(a,b)$ for two different $L = 
	L_a = L_b$ values are shown in Figure \ref{fig:flin_sections}. We see that positioning either of the 
	sampling windows in regions of rapid clock transitions decreases linearity as expected. Placing the 
	reset region as late as possible consistent with not crossing into the serial clock feedthrough is also 
	seen to be optimal. However, the resulting parameters in Figure \ref{fig:optim_CDS} (centre panel) are 
	somewhat confusing. It appears that the optimised signal window is placed very "late" - well into the 
	region where the video has stopped being flat after transferring signal charge. We will discuss a	
	possible reason for this counter-intuitive result in Section \ref{sec:subtraction}. 
	
	\begin{figure} 
		\includegraphics{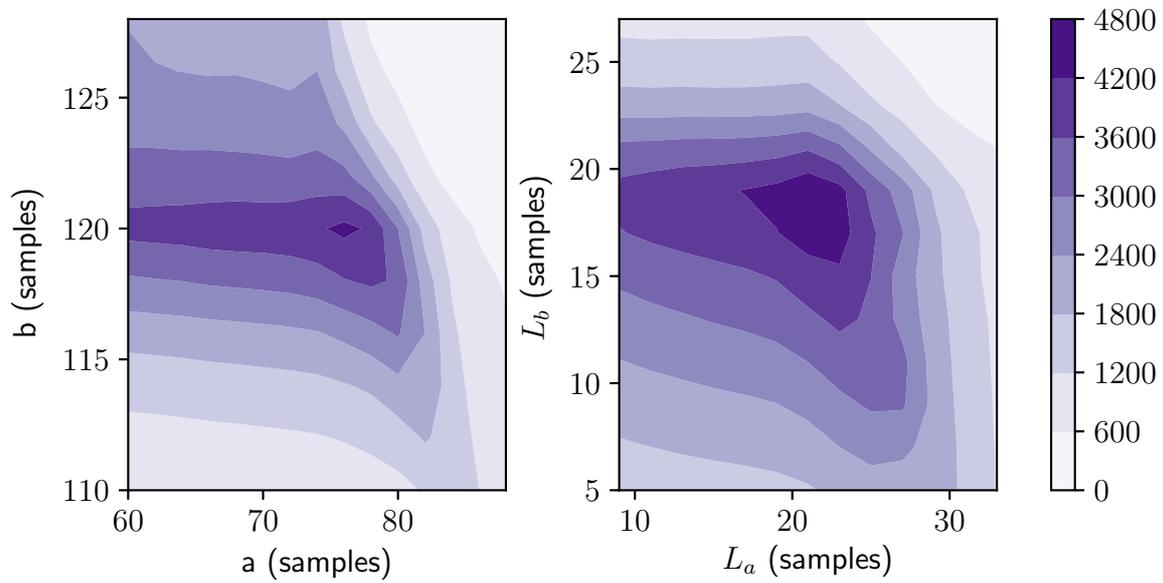}
		\caption{\label{fig:comb_sections} Cross sections for $f_\mathrm{comb}$ through $(a,b)$ at constant 
		$L_a=L_b=20$ (left panel) and through $(L_a,L_b)$ at $(a,b) = (75,120)$ (right panel)}
	\end{figure}
	Finally we show the results of a combined optimisation. Cross sections are shown in Figure 
	\ref{fig:comb_sections}, and the optimised parameters in Figure \ref{fig:optim_CDS} (right panel). The 
	chosen parameters clearly represent a compromise between the considerations of \ac{SNR} and linearity, 
	though the signal sampling window is found to be later than might be selected by eye.

\begin{figure} 
		\centering
		\includegraphics{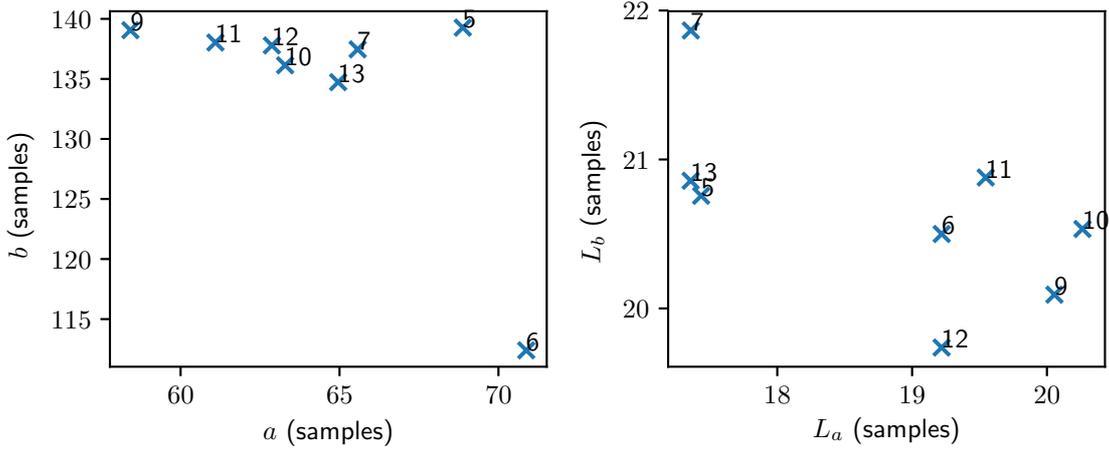}
		\caption{\label{fig:channels_optim} Combined optimised timing parameters for different \ac{CCD} 
		channels. Each point is labelled with its channel index.}
	\end{figure}

	In Figure \ref{fig:channels_optim} the parameters resulting from combined optimisation for all 8 
	channels measured are shown. A reasonably tight grouping both in terms of sample window position (except for the outlier channel 6) and window length are exhibited. It seems reasonable that selecting 
	parameter values in some centroid of the located points for various channels would be a suitable 
	compromise to optimise the entire device readout (though a higher dimensional procedure which separately 
	took into account the data for all channels simultaneously could be contemplated).

	\section{Subtraction of Raw Video Signals \label{sec:subtraction}}
	Consider the operation of a \ac{DA}, which takes a set of raw ADC samples $\hat{x}_n$ and turns them into 
	a pixel value $X_j$ with timing parameters $(a,b,L_a,L_b)$:
	\begin{equation}
		X_j = \frac{1}{L_a}\sum_{n=a}^{a+L_a}\left(\hat{x}_n\right) - \frac{1}{L_b}\sum_{i=b}^{b+L_b}\left(\hat{x}_n\right)
	\end{equation}
	
	In a scientific context, one almost universally wishes to subtract two pixel values 
	to yield a final pixel value (for example subtracting an bias frame pixel value $Y_j$ from the 
	corresponding image frame value $X_j$). Consider the output of our differential averager in such an 
	operation:
	\begin{equation} \label{multi_pix_DA}
		X_j - Y_j = \left( \frac{1}{L_a}\sum_{n=a}^{a+L_a}\left(\hat{x}_n\right) - 
		\frac{1}{L_b}\sum_{i=b}^{b+L_b}\left(\hat{x}_n\right)\right) - \left( 
		\frac{1}{L_a}\sum_{n=a}^{a+L_a}\left(\hat{y}_n\right) - 
		\frac{1}{L_b}\sum_{i=b}^{b+L_b}\left(\hat{y}_n\right)\right)
	\end{equation}
	
\begin{figure}
		\centering
		\includegraphics{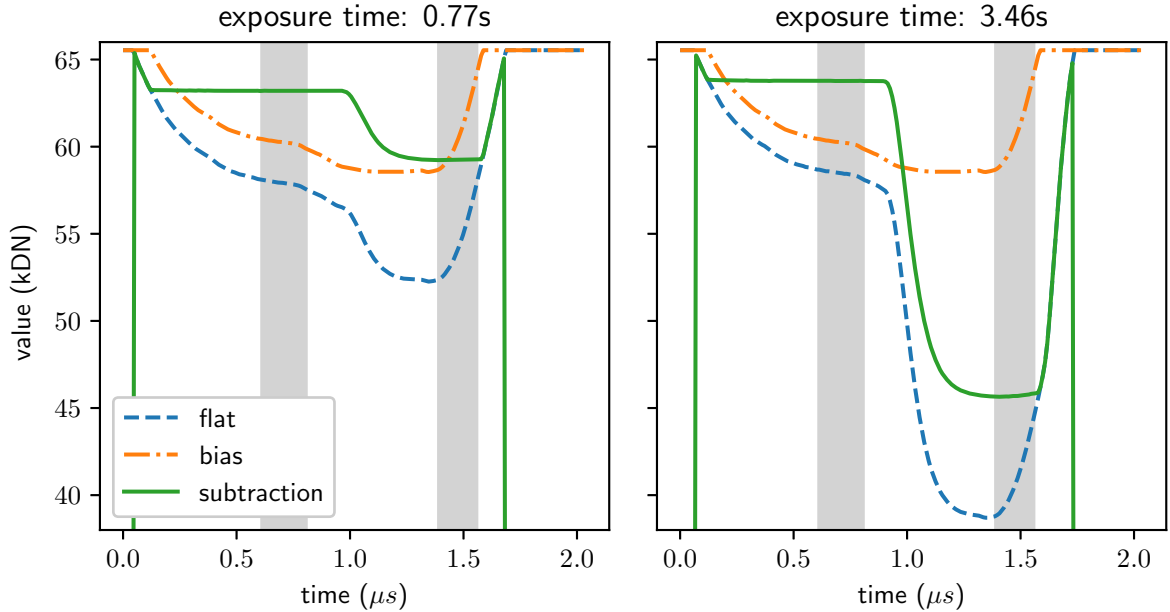}
		\caption{\label{fig:CDS_subtraction}Subtracting the raw samples of a bias pixel from an image pixel 
		for two different exposure 
		times. The shaded vertical regions show the optimised \ac{CDS} values obtained previously}
	\end{figure}

	If we can assume that the low frequency noise component and DC offset drift are small (and thus the 
	difference between the raw samples $x_n$ and $y_n$ are constituted by the response of the system to 
	different pixel values), then the linearity of the \ac{DA} operation implies we could re-arrange 
	\eqref{multi_pix_DA} to give:
	\begin{equation}
		X_j - Y_j = \frac{1}{L_a}\sum_{n=a}^{a+L_a}\left(\hat{x}_n - \hat{y}_n\right) - 
		\frac{1}{L_b}\sum_{n=b}^{b+L_b}\left(\hat{x}_n - \hat{y}_n\right)
	\end{equation}
	or, in other words: there is in principle no reason why we could not subtract the raw video for two 
	pixel values sample by sample rather than subtracting the pixel values after \ac{CDS} processing. Such an 
    operation is of very little practical use, since in reality the low frequency noise component may be significant. In addition, for acquisition of a whole image frame this method would require an excessive (and unnecessary) data 
    volume. However, this operation does provide some insight (at least in 
	the specific system readout discussed in this work) into why the optimised values of $a$ and $b$ seem to 
	be found later than would be intuited by "eyeballing" the raw samples. We show the results of this 
	operation for two randomly selected pixels in Figure~\ref{fig:CDS_subtraction}. It is much clearer 
	here how the optimised values of $a$ and $b$ might be arrived at through the procedure described in 
	Section \ref{sec:optimising} - since the rising edges at the end of the signal region are very similar 
	between the image and bias pixels, their subtraction results in the appearance of a much longer "flat" 
	portion of the pixel signal than we see in the non-subtracted traces. 
	
	We do not wish to make the claim that this effect applies to all readout systems - 
	it is likely that when $\frac{1}{f}$ noise is significant, or there is large jitter in clock 
	timings and thus inconsistent clock feedthrough positions through time, then the raw subtraction method 
	may well not give good insights into optimisation. However, in situations where these limitations do not 
	apply, this operation provides excellent insight into choosing \ac{CDS} timing values. 
	
	The raw sample subtraction method is especially intriguing in the context where a \ac{DCDS} system is not 
	available - the linearity property of the \ac{DA} operation applies equally to that of an analog circuit 
	such as a \ac{DSI}. Thus, only a few traces of pre-\ac{CDS} video data (obtained for example from an 
	oscilloscope) would be needed to get a much clearer picture for timing optimisation. 
	
	The procedure using an analog \ac{CDS} system would be roughly as follows:
	\begin{enumerate}
		\item Use the \ac{CCD} timing generator to produce a trigger pulse at some chosen pixel within the 
		image readout
		\item Use an external data acquisition system (e.g. an oscilloscope) to capture the raw video trace 
		of this pixel before \ac{CDS} processing
		\item Vary the integration time (and thus signal level) and repeat this capturing process
		\item Numerically construct raw subtracted traces for each integration time and	determine values of 
		$a$ and $b$ for which linearity performance appears acceptable.
		\item Using these $a$ and $b$ values, vary $L_a$ and $L_b$, reading out 
		whole images from the system 
		for each value, and measure the \ac{SNR} to find the maximum values consistent with the 
		required \ac{SNR}
	\end{enumerate}
	
	Unfortunately no readout system equipped with analog \ac{CDS} was available for this work. However, we 
	hope to test this optimisation procedure in future to determine its efficacy in a real world situation.
	
	\section{Conclusions}
	The effect of changing only the \ac{CDS} timing parameters with all other operating conditions of a 
	\ac{CCD} held constant has been investigated using raw sample capture from a \ac{DCDS} readout. We have 
	shown that numerical optimisation is a viable way to choose parameters which optimise a trade-off between 
	\ac{SNR} and linearity. 
	
	The observed optimal values in our particular system were observed to be counter-intuitive in the sense 
	that they appear in a region where the raw signal is rapidly changing rather than stable. We have 
	developed an explanation for this by considering the raw sample by sample subtraction of two pixel 
	values, which shows that after processing, these regions remain flat and suitable for use in \ac{CDS} 
	integration periods. No applicability of this effect is assumed for all readout systems, though the 
	insight from the subtraction method is likely to be useful in many wider contexts than considered in this 
	work. 
	
	It is recommended that in \ac{CDS} optimisation of a \ac{CCD} readout, the subtraction of two raw 
	video signals should be included as part of the inputs to choosing the timing parameters.
	
	\acknowledgments
	This work has been partially supported by STFC funding for UK participation in LSST, through grant 
	ST/N002547/1. The authors wish to acknowledge also useful conversations with Chris Damerell of Oxford 
	University.

%	\IfFileExists{references.bib}
%	    { \bibliography{references} 

\begin{thebibliography}{10}

\bibitem{1748-0221-4-03-P03002}
V.~Radeka, J.~Frank, J.~C. Geary, {\em et~al.}, ``{LSST} sensor requirements
  and characterization of the prototype {LSST} {CCD}s,'' {\em Journal of
  Instrumentation} {\bf 4}(03), P03002  (2009).

\bibitem{2008arXiv0805.2366I}
{\v Z}.~{Ivezi{\'c}}, S.~M. {Kahn}, J.~A. {Tyson}, {\em et~al.}, ``{LSST: from
  Science Drivers to Reference Design and Anticipated Data Products},'' {\em
  arXiv e-prints}   (2008).

\bibitem{janesick2001scientific}
J.~R. Janesick, {\em {Scientific charge-coupled devices}}, SPIE Press,
  Bellingham, Wash  (2001).

\bibitem{0022-3735-15-11-020}
G.~Hopkinson and D.~Lumb, ``Noise reduction techniques for {CCD} image
  sensors,'' {\em Journal of Physics E: Scientific Instruments} {\bf 15}(11),
  1214  (1982).

\bibitem{oro42476}
K.~D. Stefanov, ``Digital {CDS} for image sensors with dominant white and 1/f
  noise,'' {\em Journal of Instrumentation} {\bf 10}  (2015).
\newblock 18 pp.

\bibitem{doi:10.1117/12.2069423}
P.~R. Jorden, D.~Jordan, P.~A. Jerram, {\em et~al.}, ``e2v new {CCD} and {CMOS}
  technology developments for astronomical sensors,'' {\em Proc. SPIE} {\bf
  9154}, 91540M--91540M--15  (2014).

\bibitem{PhysRev.32.110}
H.~Nyquist, ``Thermal agitation of electric charge in conductors,'' {\em Phys.
  Rev.} {\bf 32}, 110--113  (1928).

\bibitem{1538-3873-115-811-1068}
J.~Gach, D.~Darson, C.~Guillaume, {\em et~al.}, ``A new digital {CCD} readout
  technique for ultra-low-noise {CCD}s,'' {\em Publications of the Astronomical
  Society of the Pacific} {\bf 115}(811), 1068  (2003).

\bibitem{101279}
H.~M. Wey and W.~Guggenbuhl, ``An improved correlated double sampling circuit
  for low noise charge coupled devices,'' {\em IEEE Transactions on Circuits
  and Systems} {\bf 37}, 1559--1565  (1990).

\bibitem{doi:10.1117/12.2058402}
G.~Bredthauer, ``Archon: A modern controller for high performance astronomical
  {CCD}s,''  (2014).

\bibitem{Weatherill_2017}
D.~Weatherill, K.~Arndt, R.~Plackett, {\em et~al.}, ``An electro-optical test
  system for optimising operating conditions of {CCD} sensors for {LSST},''
  {\em Journal of Instrumentation} {\bf 12}, C12019--C12019  (2017).

\bibitem{doi:10.1117/12.876627}
M.~S. Robbins, P.~Mistry, and P.~R. Jorden, ``Detailed characterisation of a
  new large area {CCD} manufactured on high resistivity silicon,'' {\em Proc.
  SPIE} {\bf 7875}, 787507--787507--12  (2011).

\bibitem{janesick2007dn}
J.~R. Janesick, {\em DN to $\lambda$}, Press Monographs, SPIE  (2007).

\bibitem{scipy}
T.~E. {Oliphant}, ``Python for scientific computing,'' {\em Computing in
  Science Engineering} {\bf 9}, 10--20  (2007).

\end{thebibliography}
%	    \bibliographystyle{spiejour}}
	    { 
 }

	%% author biographies %%
	
	\vspace{2ex}\noindent\textbf{Dan Weatherill} is a postdoctoral researcher at the University of Oxford. He 
	earned his bachelors and masters in Experimental and Theoretical Physics at the University of Cambridge, 
	and a phD in spaceflight instrumentation at The Open University. His current work mainly focuses on 
	testing, modelling and optimising the operation of CCD detectors for the LSST telescope.
	
	\vspace{2ex}\noindent\textbf{Ian Shipsey} is PI of LSST-UK CCD Characterization. He is Henry Moseley 	
	Professor of Physics at Oxford University, and a co-I of the Mu3e tracker and ATLAS HL-LHC pixel  	
	detector. He was co-PI of the CMS at LHC Phase-0 Forward pixel detector (2008), and PI of the CLEO-III 
	Silicon Vertex Detector (2000). He received the 2019 James Chadwick Medal and Prize of the UK Institute 
	of Physics for elucidation of the physics of heavy quarks.
	
	\vspace{2ex}\noindent\textbf{Richard Plackett} is the lead silicon detector scientist for the OPMD 
	cleanroom facility at Oxford University (2014); his principal interests are the upgrade for the ATLAS 
	pixel system, the Mu3e experiment, and the Medipix collaboration. He has previously worked on hybrid 
	pixel detector development for Diamond Light Source (2011), was a CERN fellow with the microelectronics 
	group (2008), and worked at Imperial College on the LHCb RICH hybrid pixel photon detector system (2002).
	
	\vspace{2ex}\noindent\textbf{Daniel Wood} is a postdoctoral researcher within the OPMD group at the 
	University of Oxford, working on silicon detector development. Prior to joining the group he completed a 
	PhD in radiation damage studies at the Centre for Electronic Imaging, the Open University.
	
	\vspace{2ex}\noindent\textbf{Kaloyan Metodiev} is a DPhil student at the University of Oxford, Particle 
	Physics Department working in the Oxford Physics Microstructure Detector laboratory (OPMD). He was born 
	in the city of Sofia, Bulgaria and completed his bachelor and masters at Trinity College, University of 
	Oxford. As part of OPMD, his main research interests lie in silicon detector devices, and he is actively 
	involved in the laboratory's work and associated projects.
	
	\vspace{2ex}\noindent\textbf{Maria Mironova} received her BSc in Physics from the University of 
	Goettingen, 
	Germany and an MSc from Imperial College London. She is now pursuing a PhD in detector development at the 
	University of Oxford.
	
	\vspace{2ex}\noindent\textbf{Daniela Bortoletto} is a professor of Particle Physics and the Head of 
	Particle Physics at the University of Oxford. She is a member of the ATLAS at CERN and the mu3e 
	collaborations at PSI. Her main research interests are Higgs physics and the development of hybrid and 
	monolithic silicon detectors for hadron and e+e- colliders. For mu3e she is focusing on  the construction 
	of a very light pixel detector.
	
	\vspace{2ex}\noindent\textbf{Nicolas Demetriou} is a postgraduate student at Imperial College London,
	interested in computational physics. He earned his Theoretical Physics degree at the University of 
	Glasgow, working on machine learning applications for beyond the standard model searches. His research
	experience also includes the investigation of birefringent materials and pressure-wave propagation in 
	granular media at German Aerospace Center, and the optimisation of CDS timing parameters at Oxford 
	Particle Microstructure Detector Facility, under the supervision Dr. Dan Weatherill.

	\listoffigures
	
\end{document}